\documentstyle[bo99,epsfig]{article}

\title{The X-ray spectra of symbiotic stars}
\author{Peter J.\  Wheatley}
\affil{Dept.\  of Physics and Astronomy,
University of Leicester, University Road, Leicester UK}

\begin{document}

\maketitle

\begin{abstract}
Symbiotic stars are thought to show distinct X-ray emission from the 
accreting object and from the colliding winds of the two
stars. I show that the colliding wind component is
unnecessary. Instead, the spectra can be interpreted as emission only
from the compact object that is strongly absorbed by the partially-ionised wind
of the red giant. There remains no evidence of any X-ray
emission from colliding winds, and thus no need for a substantial wind
from the compact object. 

\end{abstract}

\section{Introduction}
Symbiotic stars are binary stars in which usually a white dwarf accretes
from the wind of a red giant (e.g.\  Luthardt 1992). Their X-ray
spectra are apparently dominated by distinct soft and hard X-ray
components (I do not discuss the third ``supersoft''case -- usually
interpreted as steady nuclear burning of accreted material). As an
example I take the ASCA GIS2 spectrum of the bright symbiotic CH~Cyg,
plotted in Figure~1. The emission is seen to peak at 1\,keV and at
5\,keV. In Figure~1 I have also overlaid a simple 10\,keV 
bremsstrahlung spectrum that has been folded through the response of
the telescope. The dip in the observed spectrum (at
2\,keV) corresponds to a maximum in the effective area of ASCA, and
thus must be a true minimum in the X-ray emission of the CH~Cyg. 

\begin{figure}
\centerline{\psfig{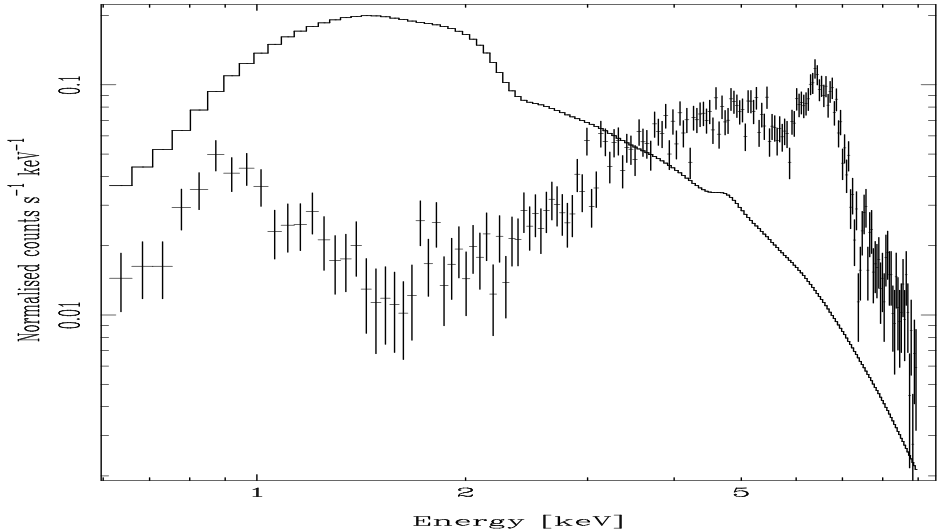}}
\caption[]{The ASCA GIS2 spectrum of the bright symbiotic star
CH~Cyg. The solid line is a simple 10\,keV bremsstrahlung spectrum
convolved with the response of ASCA, showing that the observed minimum
at $\sim$2\,keV must be a true minimum in the X-ray emission of CH~Cyg.}
\end{figure}

The ASCA observation of CH~Cyg was analysed by Ezuka, Ishida \& Makino
(1998). 
To achieve an acceptable fit they required three
emission components (kT=\\0.2, 0.7, 7.3\,keV), each with a different
absorption column, plus an additional partial-covering
absorber. Usually the hard emission is attributed to the accreting
compact object and the soft emission to colliding winds of the two
stars. In this paper I demonstrate that the spectrum can be understood
with a far more simple model, and that there is no need for a separate
soft component. 

\section{Ionised absorption}
The key to this new interpretation is to allow the absorbing medium to
be partially ionised. This
is reasonable because the wind of the red giant --- the obvious
candidate absorber --- is strongly illuminated by ionising radiation from
the accreting white dwarf. Fitting the ASCA spectrum with a
single-temperature emission model 
({\it mekal}\,)
absorbed by a photoionised medium ({\it absori}\,) I readily achieve a
fairly good fit (reduced $\chi^2$=2.4 with 172 d.o.f.; top panel of Figure~2). 
The residuals are dominated by narrow features at 0.9\,keV and
6.4\,keV. Adding narrow lines to the model I find an acceptable fit 
(reduced $\chi^2$=1.4 with 168 d.o.f.; bottom panel of Figure~2). 
Best fitting parameters are kT=11\,keV, $\rm N_H=4\times
10^{23}\,cm^{-2}$, $\xi$=840 (ionisation parameter).

\begin{figure}
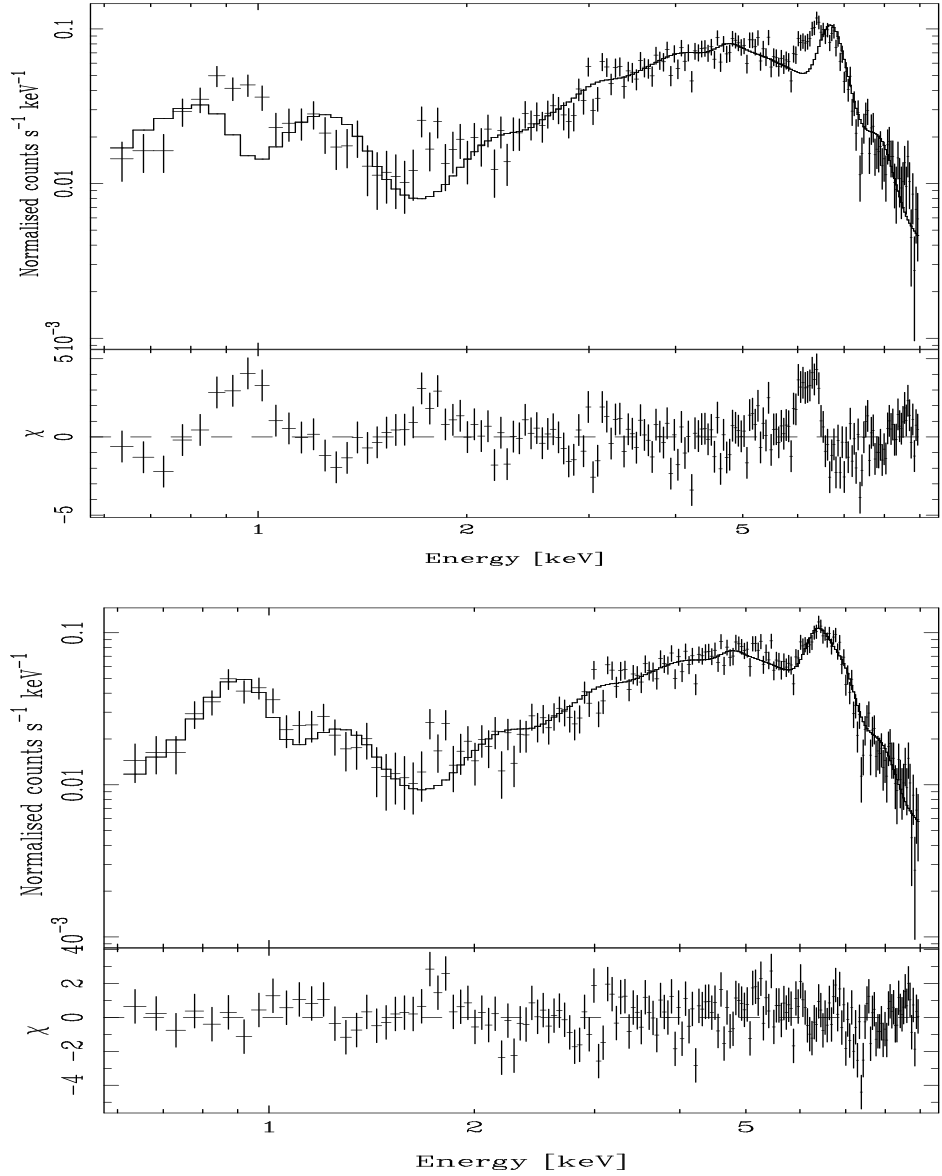

\psfig{file=fig2a.ps, width=12.4cm, height=7.5cm}

\vspace{0.5cm}

\psfig{file=fig2b.ps, width=12.4cm, height=7.5cm}
\caption[]{
The ASCA GIS2 spectrum of CH~Cyg fitted with an ionised absorption
model and a single emission component (top panel). Narrow emission
lines have been added to the model for the lower panel (at 0.9 \& 6.4\,keV), 
resulting in an acceptable fit to the spectrum 
(reduced $\chi^2$=1.4 with 168 d.o.f.).
}
\end{figure}

\section{Line emission}

The 6.4\,keV emission line is most likely due to K$_\alpha$
fluorescence of weakly-ionised iron. Since the absorbing medium is
strongly ionised, this fluorescence must arise elsewhere, probably through 
reflection from the surface of the compact object.
The 0.9\,keV line is more difficult to identify because its energy is
less well constrained and there are a large number of emission lines in
this portion of the X-ray spectrum. However, its proximity to the strong
OVIII absorption edge (see Figure~3) suggests it is most likely
the recombination continuum emission of OVIII. The emission spectrum
of the absorbing medium is neglected in the ionised absorption model 
({\it absori}\,).

\section{No need for colliding winds}

The consequence of my fit to the ASCA spectrum of CH~Cyg is that a
separate low-temperature emission component is no longer needed.
Thus there is no need to invoke X-ray emission from colliding winds in
this system, 
and indeed, no longer any need for a substantial wind from the white dwarf.

The model spectrum (Figure~3) shows how the partially ionised absorber
cuts deeply at intermediate energies but allows soft photons to leak
through. Most evidence taken to support X-ray emission from colliding
winds has come from soft X-ray observations, e.g.\  ROSAT (M\"{u}rset,
Wolff \& Jordan 1997). Clearly the ROSAT spectrum (0.1-2.5\,keV) of an
absorbed system will reveal only the soft X-ray leak, and this could
be mistaken for a low temperature emission spectrum. I believe that
all the ROSAT spectra of symbiotic stars previously interpreted as
emission from colliding winds may be reinterpreted as absorbed hard
X-ray spectra. 

\begin{figure}
\centerline{
\psfig{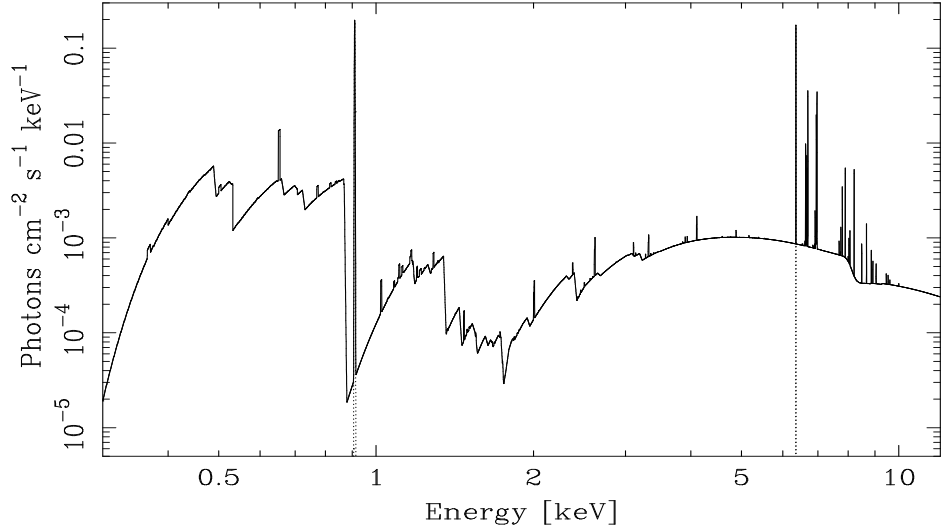}
}
\caption[]{The model spectrum fitted to the ASCA spectrum in the lower
panel of Figure~2. }
\end{figure}


\begin{references}
\ref Luthardt, R. 1992, RvMA 5, 38

\ref Ezuka, H., Ishida, M. \& Makino, F. 1998, ApJ 499, 388 

\ref M\"{u}rset, U., Wolff, B. \& Jordan, S. 1997, A\&A 319, 201

\end{references}
\end{document}